\documentclass [twocolumn, aps, showpacs] {revtex4}
\usepackage{graphicx}
\begin{document}
\draft
\title {Resistance due to vortex motion in the $\nu=1$ bilayer quantum Hall superfluid}
\author {David A. Huse}
\address{Department of Physics, Princeton
University, Princeton, NJ 08544}
\date{\today}
\begin{abstract}
The longitudinal and Hall resistances
have recently been measured for quantum Hall bilayers at total
filling $\nu=1$ in the superfluid state with interlayer pairing,
both for currents flowing parallel to one another and for
``counterflowing" currents in the two layers.  Here I examine the
contribution to these resistances from the motion of unpaired
vortices in these systems, developing some possible explanations
of various qualitative features of these data.
\end{abstract}
 \maketitle

The $\nu=1$ interlayer superfluid quantum Hall state occurs in
systems where carriers (electrons or holes) are confined to move
two-dimensionally in two closely-spaced parallel layers (quantum
wells), subject to a perpendicular quantizing magnetic field of near
one flux quantum per carrier \cite{wen, yang, moon, kell02, kellogg,
tutuc, sheng}. It occurs when the tunneling between the layers is
negligible, but the layers are close enough together relative to the
spacing between carriers that interlayer correlation due to the
Coulomb repulsion is strong.  The bosons that condense to make a
superfluid here are pairs consisting of an electron in one layer and
a hole in the other, so they have zero net charge.  Recent
experiments \cite{kellogg, tutuc} have looked explicitly at this
superfluidity by contacting separately to each layer to produce a
current of these interlayer dipoles: counterflowing electrical
currents of equal magnitude but opposite direction in the two
layers.  What is found is that these systems are apparently
superfluid only in the zero temperature limit.  At nonzero
temperature, dissipation is seen, as a nonzero longitudinal
resistance, $R_{xx}$, and this dissipation is found to be a little
larger for counterflowing currents as compared to parallel currents
that are identically directed in the two layers \cite{kellogg,
tutuc}. As discussed below, this, and also some features of the
counterflow Hall resistance, can possibly be understood
phenomenologically in terms of the motion of the vortices of this
superfluid. Here I will consider only ``balanced'' bilayers, where
the average carrier density is the same in each of the two layers,
although the behavior as these bilayers are imbalanced is also
interesting \cite{spiel}.

The elementary vortices of this superfluid carry a quantized charge
of $\pm e/2$ in addition to their quantized vorticity \cite{moon}.
Thus there are four types of vortices, with charge and vorticity of
either sign.  In order to specify these signs unambiguously, we need
to set some sign conventions:  The carriers are confined to layers
parallel to the $xy$ plane, and the perpendicular component of the
magnetic field points along the positive $z$ direction.  The ``top"
layer is the layer at larger $z$.  The vector representing a
counterflowing current density points in the direction of the
electrical current in the top layer and opposite to that in the
bottom layer.  A positive interlayer electric dipole has positive
charge in the top layer and negative charge in the bottom layer.  A
vortex with positive vorticity has its vorticity (as set by a
right-hand rule on the circulating counterflowing currents) pointing
in the positive $z$ direction: viewed from above, this vortex's
electrical currents flow counterclockwise in the top layer and
clockwise in the bottom layer.  In addition to its charge and
vorticity, the elementary vortices also carry an unquantized
interlayer electric dipole \cite{moon}.  With the above sign
conventions, the sign of a vortex's dipole is given by the product
of the signs of its charge and vorticity.  This is demonstrated,
using approximate wavefunctions \cite{moon} of these vortices, in
the final few paragraphs of the present paper.

In an {\it ideal} sample with no randomness and exactly at total
filling $\nu=1$, the ground state has no vortices.  The vortices in
such a two-dimensional superfluid of interlayer dipoles interact
logarithmically at large distances, and remain bound in pairs with
zero total vorticity up to a nonzero Kosterlitz-Thouless (KT)
transition temperature \cite{wen,yang,moon}.  [Note, the
vortex-vortex interaction energies due to the electric charges and
dipoles on the vortices fall off with distance, so it is the
logarithmic interaction due to the vorticity and the superfluidity
that dominates their interaction at large distance.]  Below the KT
transition temperature a counterflowing current (which is a
supercurrent of dipoles) flows without linear-response resistance in
such an ideal sample. However, this is not what is seen in the
recent experiments \cite{kellogg, tutuc}, where the superfluidity
(zero resistance) is observed to be present only in the zero
temperature limit.  This is likely due to random potential disorder
in the samples studied experimentally.  A random potential couples
to both the electric charge and the dipole moment of the vortices
and thus if strong enough can stabilize a ground state with
unpaired, pinned vortices in a pattern specific to the particular
random potential in each sample.  [One can see that the disorder is
in some sense rather strong in both of the experimental samples
\cite{kellogg, tutuc} by noting that they enter the insulating phase
at magnetic fields just above those that produce the $\nu=1$ state
we are discussing here.] Such a ground state is a type of
vortex-glass \cite{sheng} that is superfluid at zero temperature but
not at any positive temperature. This appears to be the situation in
the recent experiments \cite{kellogg, tutuc}. In such a vortex glass
state it is the thermally-activated motion of these pinned, unpaired
vortices that should dominate the low temperature resistance. Thus
it seems worthwhile to look more closely at the motion of these
vortices in response to applied currents.

First, let's look at the {\it forces} on the vortices due to the
currents, at low temperature.  Let ${\bf J}$ be the current
density {\it per layer}. For {\it parallel} currents, where ${\bf
J}$ is the same in both layers, there is a Hall electric field of
magnitude $E=2Jh/e^2$ perpendicular to the current, since we are
in a $\nu=1$ quantum Hall state and the total current density is
$2{\bf J}$.  This Hall electric field couples to the charge $\pm
e/2$ of the vortices, producing a force on each vortex that is
perpendicular to the current and is of magnitude
\begin{equation}
F_v=Jh/e~.
\end{equation}
Now for {\it counterflowing} electrical currents, on the other
hand, we have a supercurrent density of the interlayer
electron-hole pairs (they are bosons) of ${\bf J}_d={\bf J}/e$
(net number of pairs per time per length), where $\pm{\bf J}$ is
the electrical current in the top/bottom layer. As is standard in
superfluidity, this supercurrent interacts with the vorticity,
producing a Magnus force on the elementary vortices that is
perpendicular to the current and of magnitude $F_v=hJ_d=Jh/e$.
Thus we find that at this level of approximation the force on a
vortex due to a current is of the same magnitude for parallel and
counterflowing currents.  When the vortices move in a dissipative
fashion along the direction of these forces, this produces
electrical resistance. The equality of the magnitude of the forces
indicates why the longitudinal resistance, $R_{xx}$, is found to
be of (roughly) similar magnitude for both types of currents in
the recent experiments \cite{kellogg, tutuc}.  However, although
they are of similar magnitude, $R_{xx}$ in fact is measured to be
somewhat larger for the counterflowing currents than for the
parallel currents in both experiments \cite{kellogg, tutuc}; I
next explore some possible reasons for this difference.

When a vortex moves, there are (at least) two effects in addition
to the forces discussed above that might enter.  First, the vortex
may tend to drift {\it along} (or opposite to) the current.  And,
when it moves it is also subject to the Lorentz force due to its
charge moving through the magnetic field.  Here I will use the
term ``drift" for the component of a vortex's motion parallel to
the current.  I make what appear to be reasonable assumptions
about the sign of the direction of this drift.  However, better
(more microscopic?) arguments to support (or counter) these
assumptions would be desirable.

For {\it parallel currents} the dissipation is reduced if the
vortices tend to drift in the same direction as the carriers'
motion.  The Lorentz force due to such vortex drift with the
carriers opposes the force on the vortex due to the Hall voltage. If
the vortex were to drift along at the same speed as the carriers,
the net force on it would vanish, just as it does for the carriers.
But since the vortices are pinned, the expectation is that their
drift speed is less than that of the carriers.  The reduction in the
force on the vortices due to their drift presumably reduces the rate
at which they hop or tunnel in the direction perpendicular to the
current, and thus a drift effect of this sign {\it reduces} the
dissipation from what it would be if the vortices move only
perpendicular to the current.
It is not clear how one would be able to detect experimentally to
what extent the vortices are drifting along the current in this case
of parallel currents.

For {\it counterflowing currents} the carriers are moving in
opposite directions in the two layers, so the current by itself
cannot dictate the direction in which a vortex drifts. However, this
is a current of electric dipoles and each vortex does carry a dipole
moment that can determine the direction the vortex drifts. Another
way of viewing it is that the density of carriers is imbalanced at
the core of each vortex (thus it has an electric dipole), and the
direction in which a vortex drifts is the same as the direction of
carrier motion in the layer where the vortex has more carriers.
Again, there is a Lorentz force due to the motion of this charged
vortex, but in this case it adds to the Magnus force due to the
supercurrent, increasing the dissipation. For example, let's
consider the case of a vortex with positive charge, dipole and
vorticity with the current of dipoles along the positive $x$
direction.  In this case the Magnus force is along the negative $y$
direction (its direction is set by the current and the vorticity).
When this vortex drifts along the positive $x$ direction (due to its
positive dipole), the resulting Lorentz force is also along the
negative $y$ direction and adds to the Magnus force. The direction
of the Lorentz force is dictated by the current, charge and dipole.
For all vortices, the sign of their charge times that of their
dipole is equal to the sign of their vorticity, so this addition of
the two forces and consequent increase in the dissipation occurs for
all four vortex types.

Thus we see that the experimental observation that $R_{xx}$ is
larger for counterflowing rather than parallel currents may be due
to a tendency of the vortices to drift along with the carriers in
a parallel current and/or a tendency of the vortices to drift in
the same direction as pairs with the same sign dipole in the case
of a counterflowing current.

The motion of the vortices along the current also gives a
contribution to the Hall resistance.  For a parallel current, the
Hall resistance is large, and the small current carried by the
charge of the moving vortices is negligible at low temperature
compared to the total current.  Thus it appears that the
contribution of the vortices to the Hall resistance will be too
small a correction to detect for a parallel current.

For a {\it counterflowing current}, on the other hand, the Hall
resistance vanishes in the low temperature limit, and vortex
motion along the current should give a significant contribution to
the Hall resistance at low $T$.  What I find is that the sign of a
vortex's contribution to the counterflow Hall resistance is given
by its charge:  For example, let's look again at the vortex with
positive charge, dipole and vorticity, with a current of dipoles
along the positive $x$ direction.  This vortex moves along the
negative $y$ direction due to the Magnus and Lorentz forces on it,
and it drifts with the current, along the positive $x$ direction.
The electric field due to this vortex motion through the
superfluid is perpendicular to the vortex motion and thus has a
Hall component along the positive $y$ direction in the top layer.
This contribution to the Hall resistance is of the same sign as
the conventional Hall resistance of positively charged carriers
(holes). The sign of this contribution to the Hall resistance is
set by the product of the sign of the vortex's dipole, which
determines the direction of the vortex's motion along the current,
and the sign of its vorticity, which sets the sign of the
resulting electric field. But this product is the sign of the
vortex's charge.  So, to summarize, the sign of the contribution
to the counterflow Hall resistance due the motion of a vortex
along the current is the same as that of the conventional Hall
resistance for carriers with the same sign charge as the vortex.

When we refer to the net charge of a vortex, this means the
difference in charge from a uniform $\nu=1$ state.  Thus for
filling $\nu=1$, there must be an equal number of vortices of
positive and negative charge.  However, there is in general no
particle-hole symmetry, so the core energies, mobilities, and
tendencies to drift need not be equal for positively and
negatively charge vortices.  This allows the sign of the total
vortex contribution to the low-temperature counterflow Hall
resistance to be set by the specific particle-hole asymmetries of
the system. In the experiment on holes \cite{tutuc} the
counterflow Hall angle is near zero at low temperature, suggesting
that the contributions from the two signs of vortex charge are
similar in magnitude and (almost) cancel in this sample.  In the
data on electrons \cite{kellogg}, on the other hand, the
counterflow Hall angle at $\nu=1$ appears to remain nonzero and of
the same sign as the conventional Hall effect for electrons,
suggesting that the negatively charged vortices dominate in this
sample, perhaps due to higher mobility and/or higher tendency to
drift along the current.

Next, let's consider the behavior as we move away from total
filling $\nu=1$.  If we move well away, the quantum Hall effect is
lost, the longitudinal resistance becomes large, the interlayer
pairing is lost and the counterflow Hall resistance becomes large
and similar in magnitude to the Hall resistance for a parallel
current.  These same things also happen as the temperature is
raised.  Thus there is a general trend as one moves away from
$\nu=1$ and $T=0$ for the counterflow Hall resistance to increase
in magnitude to near the conventional value (and sign) for the
given density of carriers in each individual layer.  At low
temperature closer to $\nu=1$ there is another trend that appears
to be in the data \cite{kellogg, tutuc, tut2} and may be due to
the vortex motion. For filling larger than but near $\nu=1$, there
are more carriers than flux quanta, and the excess charge will sit
on the vortices, so there are now more vortices with the same sign
charge as the carriers than there are with the opposite charge.
These vortices give a contribution to the counterflow Hall angle
of the same sign as the system has well away from $\nu=1$. For
fillings less than $\nu=1$, the vortices that give the opposite
sign contribution are more prevalent.  Thus we expect that the
counterflow Hall angle will increase towards its ``normal'' value
as $\nu$ is increased from $\nu=1$, but as $\nu$ is decreased the
counterflow Hall may first decrease due to the polarization of the
vortices' charge before it increases due to the disruption of the
pairing.  It seems better to use the counterflow Hall {\it angle}
data to look for this effect, since the individual counterflow
resistances both have strong and somewhat similar dependences on
$\nu$ and $T$. Converting them together into the counterflow Hall
angle removes some of this strong dependence.  Such an asymmetry
of the counterflow Hall angle about $\nu=1$ is indeed there for
the hole samples \cite{tut2}, and appears to be there in Fig. 2b
of \cite{kellogg}.

Finally, let's look at approximate wavefunctions for the
superfluid ground state and its vortices in an ideal,
disorder-free bilayer, to determine the signs of the charge,
dipole and vorticity of each vortex.  Here I follow the paper of
Moon, {\it et al.} \cite{moon}. We work in the lowest Landau level
(LLL), using Coulomb gauge and the orbitals with angular momentum
$m=0,1,...$ about the origin. Let $c_m^{\dagger}$ create a carrier
in the top layer in the LLL in orbital $m$, while $b_m^{\dagger}$
creates a carrier in the same orbital in the bottom layer.  Then
the ground state of the superfluid is, to first approximation in
the inter-carrier Coulomb interaction,
\begin{equation}
|\Psi_0\rangle=\prod_{m\geq
0}{{1}\over{\sqrt{2}}}(c_m^{\dagger}+b_m^{\dagger})|0\rangle~,
\end{equation}
where $|0\rangle$ is the ``vacuum" of no carriers.  In this
wavefunction, whenever orbital $m$ is occupied in the top layer,
it is empty in the bottom layer, and vice versa.  Thus it has
interlayer electron-hole pairing.  To minimize the interaction
energy, the average occupancy of the two layers is equal, and the
relative phase between the amplitudes for the carrier being in the
two layers is spatially uniform in order to minimize the exchange
energy.

To make one type of vortex, instead pair orbital $m$ in the bottom
layer with $m+1$ in the top layer:
\begin{equation}
|\Psi_v\rangle=\prod_{m\geq
0}{{1}\over{\sqrt{2}}}(b_m^{\dagger}+c_{m+1}^{\dagger})|0\rangle~.
\end{equation}
This vortex state has on average 1/2 of a carrier missing in the
top layer at the center of the vortex, but the same average
density as the ground state in the bottom layer.  Since the
missing charge is only in the top layer, the net charge and the
dipole moment of this vortex have the same sign (in fact, they
both have the opposite sign from the charge of the carriers). To
zero-th order in the inter-carrier Coulomb interaction this vortex
does not have circulating currents, since LLL states in the
absence of a potential energy do not carry net current. The
superfluid density, and thus the currents, are due to the
interactions \cite{moon}, so to get the sign of the current we
must examine the change of this vortex wavefunction due to the
interactions, for example in a Hartree-Fock approximation.

Consider the single-particle state in our vortex that is a linear
combination of orbital $m$ in the bottom layer and $m+1$ in the
top layer.  Orbital $m$ is concentrated a little closer to the
center of the vortex than $m+1$.  This difference in guiding
center radius is proportional to $1/\sqrt{m}$, while the radius
itself is proportional to $\sqrt{m}$.  This carrier is repelled by
all the other occupied states in both layers, so the effective
potential it sees has a minimum at a radius somewhere between
these two orbitals' centers.  This potential (which produces an
attraction between the electron and hole) perturbs the two
orbitals so that orbital $m$ in the bottom layer is displaced
outwards and as a result has a diamagnetic net current, while
orbital $m+1$ in the top layer is displaced inwards and has a
paramagnetic net current. Thus we find that for this vortex the
currents are paramagnetic in the top layer, and thus, by the
right-hand rule convention I am using, the vorticity is positive,
and the product of the signs of the vortex's three attributes,
charge, dipole and vorticity, is positive.  If we instead pair $m$
in the top layer with $m+1$ in the bottom layer, this reverses the
dipole and the vorticity, but leaves the net charge unchanged, so
the product of the three signs remains positive.  To change the
vortex's net charge, the empty $m=0$ state is filled with a
carrier \cite{moon}: this reverses the signs of the charge and the
dipole, but leaves the vorticity unchanged.

I thank Emanuel Tutuc and Mansour Shayegan for numerous
discussions about their experiments, Steve Girvin for discussion
about these vortices, and the NSF for support through MRSEC grant
DMR-0213706.

\end{document}